\documentclass[11pt]{article}
\usepackage{epsfig} 
\usepackage{amssymb} 
\usepackage{graphics}
\newcommand{\sect}[1]{ \section{#1} \setcounter{equation}{0} }

\newcommand{\half}{\mbox{\small{$\frac{1}{2}$}}}

\newcommand{\MSbar}{\overline{\mbox{MS}}} 
 
\newcommand{\Cc}{\mathbb{C}}
\newcommand{\Pp}{\mathbb{P}}

\begin{document}

\title{Six dimensional ultraviolet completion of the $\Cc\Pp(N)$ $\sigma$ model
at two loops}
\author{J.A. Gracey, \\ Theoretical Physics Division, \\
Department of Mathematical Sciences, \\ University of Liverpool, \\ P.O. Box
147, \\ Liverpool, \\ L69 3BX, \\ United Kingdom.}
\date{}

\maketitle

\vspace{5cm}
\noindent
{\bf Abstract.} We extend the recent one loop analysis of the ultraviolet
completion of the $\Cc\Pp(N)$ nonlinear $\sigma$ model in six dimensions to two
loop order in the $\MSbar$ scheme for an arbitrary covariant gauge. In 
particular we compute the anomalous dimensions of the fields and 
$\beta$-functions of the four coupling constants. We note that like Quantum 
Electrodynamics (QED) in four dimensions the matter field anomalous dimension 
only depends on the gauge parameter at one loop. As a non-trivial check we 
verify that the critical exponents derived from these renormalization group 
functions at the Wilson-Fisher fixed point are consistent with the $\epsilon$ 
expansion of the respective large $N$ exponents of the underlying universal 
theory. Using the Ward-Takahashi identity we deduce the {\em three} loop 
$\MSbar$ renormalization group functions for the six dimensional ultraviolet 
completeness of scalar QED.

\vspace{-18.5cm}
\hspace{13.2cm}
{\bf LTH 1232}

\newpage

\sect{Introduction.}

There has been wide interest in recent years in studying the ultraviolet
completion of quantum field theories beyond their critical dimension. For
instance, $O(N)$ $\phi^4$ theory, which is renormalizable in four dimensions,
has been completed to six dimensions and is related to $O(N)$ $\phi^3$ theory.
This has been verified in detail in \cite{1,2} as well as for other related
field theories, \cite{3}. Briefly a common core interaction between the matter 
and force fields present in both theories is responsible for the dynamics at 
the Wilson-Fisher critical point in $d$-dimensions, \cite{4}. Through that 
interaction the canonical dimensions of the fields are defined and thereby 
determine the relevant operators of the respective theories in their critical
dimensions. In other words there is a universal theory built with an infinite
number of operators constructed from all the fields, \cite{5}, a finite subset
of which are relevant in successive critical dimensions. Accessing the 
properties of this theory allows one to connect $O(N)$ $\phi^4$ and $\phi^3$ 
theories in their respective critical dimensions. One calculational tool to 
achieve this is the large $N$ expansion which provides the $d$-dimensional 
critical exponents available at several orders in the parameter $1/N$ through 
the pioneering papers \cite{6,7,8}. This parameter acts as a dimensionless
perturbative coupling constant in all dimensions in this limit. Expanding such 
exponents in an $\epsilon$ expansion about each critical dimension the 
coefficients of the Taylor series are in one-to-one agreement with the 
$\epsilon$ expansion of the renormalization group functions of the respective 
theories at each of their Wilson-Fisher fixed points. Indeed the seven loop 
$O(N)$ $\phi^4$ renormalization group functions, 
\cite{9,10,11,12,13,14,15,16,17,18}, and those of $\phi^3$ theory at four 
loops, \cite{1,2,19,20,21,22}, have been shown to be in precise agreement with
the exponents of \cite{6,7,8}.

Having confirmed this connection through high order computations for a 
well-studied set of scalar theories, other universality classes have 
subsequently been probed. Recently this has been undertaken for another class, 
similar to the scalar case already mentioned, which is that of the nonlinear 
$\Cc\Pp(N)$ $\sigma$ model, \cite{23,24}. The critical dimension of this field 
theory is two and the model is parallel to the $O(N)$ nonlinear $\sigma$ model 
which serves as the base theory in the tower of theories that includes $O(N)$ 
$\phi^4$ and $\phi^3$ theory. In even dimensions one can construct a 
renormalizable Lagrangian with the same core symmetries as the base two 
dimensional theory so that each of these higher dimensional models is a member 
of the same tower. In the $\Cc\Pp(N)$ case, as the scalar fields are complex 
conjugates, a $U(1)$ spin-$1$ field is also present in two dimensions in 
addition to the model's eponymous spin-$0$ scalar field. This $U(1)$ spin-$1$ 
field of two dimensions becomes a gauge field in the tower of theories above 
two dimensions. Both the scalar $\sigma$ and $U(1)$ fields, which we notionally
regard as force fields, couple to the complex matter fields and it is these two
interactions that drive the critical point dynamics in the $d$-dimensional 
universal theory. We note that when the coupling constant of the $\sigma$ field
to matter is formally switched off the universality class corresponds to that 
of scalar QED, \cite{23,24,25,26,27,28}. In \cite{23,24} the ultraviolet 
completion to six dimensions was considered and a comprehensive Landau gauge 
one loop computation of renormalization group functions was carried out. Unlike
the $O(N)$ counterpart the six dimensional extended $\Cc\Pp(N)$ $\sigma$ model
has four interactions. However the resulting renormalization group functions 
were shown to be consistent with the large $N$ critical exponents of the 
underlying universal theory computed in \cite{28,29}. 

Given the establishment of this class and its perturbative analysis at one 
loop, it is the purpose of this article to extend the renormalization group 
functions of the six dimensional theory to two loop order. This is not a 
trivial task. For instance, of necessity when constructing the ultraviolet
completion the force fields have propagators that have an additional power
of the momentum in the momentum space representation. In the case of the $U(1)$
gauge field this means that it has a dipole propagator structure. Therefore
this complicates the evaluation of all the two loop Feynman graphs that need
to be computed. Therefore we had to appeal to various modern ways of carrying
out the renormalization. In computing the anomalous dimensions and
$\beta$-functions in an arbitrary linear covariant gauge we will establish the
connection to the exponents of the universal theory at a new loop order. En
route we will partially check the result by carrying out the {\em three} loop
field anomalous dimensions. An interesting corollary to this is that we will
be able to deduce the full three loop renormalization group functions of the
ultraviolet completion of scalar QED in six dimensions. This follows trivially
since the $\beta$-function of the gauge field, which is the only core coupling 
constant in this class in six dimensions and drives the critical dynamics, can 
be deduced from the Ward-Takahashi identity similar to the one widely known in 
standard fermionic QED.

The article is organized as follows. The background to the $\Cc\Pp(N)$ 
universality class and the six dimensional Lagrangian are briefly reviewed in
the next section. Subsequently section 3 is devoted to recording the results
for the renormalization group functions including discussion on the various
checks undertaken to ensure their credibility. This includes reconciling the
$\epsilon$-expansion of the critical exponents with their known large $N$  
counterparts. We present concluding remarks in section 4.

\sect{Background.}

Briefly the background to the universality class which includes scalar QED
begins with the two dimensional theory which serves as the foundation for the
tower we consider here. In order to have a conserved charge one has to have 
complex scalar fields and that determines the two dimensional theory to be the 
$\Cc\Pp(N)$ nonlinear $\sigma$ model which has the Lagrangian
\begin{equation}
L^{(2)} ~=~ \overline{D_\mu \phi}^{\,i} D^\mu \phi^i ~+~ 
\sigma \left( \bar{\phi}^i \phi^i ~-~ \frac{1}{g_2} \right)
\label{lag2}
\end{equation}
where $1$~$\leq$~$i$~$\leq$~$N$ and $D_\mu$ is the usual covariant derivative
involving the field $A_\mu$. At this stage we do not refer to it as a gauge
field since on dimensional grounds it has no kinetic term and therefore
corresponds to an auxiliary field. We will always denote the coupling constant
ordinarily associated with the gauge field appearing in the covariant 
derivative by $g_1$ for reasons that will become clear later. The remaining
coupling constant $g_2$ has been scaled out of the interaction involving the
scalar field $\sigma$ since it is this interaction as well as the cubic one of
the $\phi^i$ kinetic term that drives the universality class across all
dimensions via the Wilson-Fisher fixed point. In other words the canonical
dimensions of $\sigma$ and $A_\mu$ are respectively $2$ and $1$ in 
$d$-dimensions given that $\phi^i$ has canonical dimension $(\half d$~$-$~$1)$.
Thus all the terms in (\ref{lag2}) have the same dimension and the theory is 
renormalizable. 

With these canonical scaling dimensions and the universal interactions of the
universality class the Lagrangians for the theories in the same class that are
renormalizable in higher dimensions are straightforward to write down. The
method is to construct all possible independent interaction terms consistent
with the critical dimension of spacetime dimension of interest and associate
separate coupling constants with each. The only caveat is that one must ensure
that the construction is consistent with the underlying symmetries. In this 
case these are the $U(1)$ symmetry due to the complex scalar and the 
$\Cc\Pp(N)$ symmetry. In addition the former symmetry now becomes a gauge 
symmetry beyond two dimensions. Therefore in four dimensions the next theory in
the tower has the Lagrangian 
\begin{equation}
L^{(4)} ~=~ \overline{D_\mu \phi}^{\,i} D^\mu \phi^i ~+~ 
\frac{1}{2} \sigma^2 ~-~ \frac{1}{4} F_{\mu\nu} F^{\mu\nu} ~-~ 
\frac{1}{2\alpha} \left( \partial^\mu A_\mu \right)^2 ~+~ 
g_2 \sigma \bar{\phi}^i \phi^i  
\label{lag4}
\end{equation}
where $F_{\mu\nu}$~$=$~$\partial_\mu A_\nu$~$-$~$\partial_\nu A_\mu$ and 
$\alpha$ is the gauge parameter. In another sense one can regard $\alpha$ as a
coupling constant of a $2$-point interaction. Setting $g_2$~$=$~$0$ corresponds
to scalar QED. Repeating the argument but with six as the critical dimension 
one arrives at the renormalizable Lagrangian 
\begin{eqnarray}
L^{(6)} &=& \overline{D_\mu \phi}^{\,i} D^\mu \phi^i ~+~ 
\frac{1}{2} \partial^\mu \sigma \partial_\mu \sigma ~-~ 
\frac{1}{4} \partial_\mu F_{\nu\sigma} \partial^\mu F^{\nu\sigma} ~-~ 
\frac{1}{2\alpha} \left( \partial_\mu \partial^\nu A_\nu \right)
\left( \partial^\mu \partial^\sigma A_\sigma \right) \nonumber \\
&& +~ g_2 \sigma \bar{\phi}^i \phi^i ~+~ \frac{g_3}{6} \sigma^3 ~+~
\frac{g_4}{2} \sigma F_{\mu\nu} F^{\mu\nu}
\label{lag6}
\end{eqnarray}
which was first given in \cite{23} for the next Lagrangian in the $\Cc\Pp(N)$ 
tower of theories. In this dimension the $\sigma$ field becomes propagating for
the first time and the gauge condition remains as the usual Lorenz one with 
$\partial^\mu A_\mu$~$=$~$0$ but contained in the Lagrangian in a dimensionally
consistent way. We note that we have defined our new coupling constants 
differently to \cite{23} and more in keeping with previous work, \cite{29}. 
Indeed like \cite{29} the gauge field has a double pole propagator that has 
also been studied in a more general six dimensional gauge theory in \cite{30}.

\sect{Results.}

Having reviewed the context in which the six dimensional extension of the
$\Cc\Pp(N)$ $\sigma$ model sits in the tower of theories of the universality 
class we now turn to establishing this at the two loop level by explicit 
computation of all the renormalization group functions. The method we have 
followed to achieve this has been documented in \cite{22,29} and we refer the 
reader to those articles for technical details. Though we note that to 
determine the $\beta$-functions we had to compute each of the $3$-point 
functions for the off-shell symmetric point configuration. By contrast in 
\cite{22} the four loop renormalization group functions of scalar $\phi^3$ 
theory in six dimensions were determined by solely considering $2$-point 
functions. In that case the $3$-point functions that needed to be renormalized 
were generated by a simple mapping of the propagator that was infrared safe. 
While it appears that the same technique could be applied to (\ref{lag6}) due 
to the cubic interactions, it is not possible since there is a quartic 
interaction in addition. A contribution from such a vertex cannot be generated 
from the mapping construction given in \cite{22} which is the reason why we 
have had to compute the two loop vertex functions directly. Finally we note 
that all our computations used the Laporta algorithm, \cite{31}, and 
specifically its {\sc Reduze} encoding, \cite{32}. The overall computation was 
carried out automatically using the symbolic manipulation language {\sc Form}, 
\cite{33,34}, where (\ref{lag6}) was dimensionally regularized in 
$d$~$=$~$6$~$-$~$2\epsilon$ dimensions. The Feynman diagrams were generated 
with {\sc Qgraf}, \cite{35}. For example there were $155$, $122$, $94$ and 
$122$ two loop graphs for the $3$-point vertex functions associated with $g_1$ 
to $g_4$ respectively.

Having outlined the method of computation we now present our results. First the
renormalization group functions of the fields are
\begin{eqnarray}
\gamma_A(g_i) &=& 
-~ \frac{N g_1^2}{30} 
\nonumber \\
&& +~ \frac{1}{1080} \left[ 10 N g_1^2 g_2^2 - 370 N g_1^4 
+ 6 N g_1^2 g_4^2 + 30 N g_2^2 g_4^2 + 15 g_3^2 g_4^2 - 120 g_3 g_4^3
+ 660 g_4^4 \right] \nonumber \\
&& +~ \frac{1}{1944000}
\left[ 
358000 N g_1^4 g_2^2 
- 40680 N^2 g_1^6 
- 2384000 N g_1^6 
- 15000 N^2 g_1^4 g_2^2 
\right. \nonumber \\
&& ~~~~~~~~~~~~~~~~ \left.
-~ 2130 N^2 g_1^4 g_2 g_4 
- 537000 N g_1^4 g_2 g_4 
+ 354 N^2 g_1^4 g_4^2 
\right. \nonumber \\
&& ~~~~~~~~~~~~~~~~ \left.
-~ 207000 N g_1^4 g_4^2 
- 10500 N^2 g_1^2 g_2^4 
+ 16000 N g_1^2 g_2^4 
+ 28500 N g_1^2 g_2^3 g_3 
\right. \nonumber \\
&& ~~~~~~~~~~~~~~~~ \left.
-~ 5250 N^2 g_1^2 g_2^3 g_4 
+ 36000 N g_1^2 g_2^3 g_4 
- 5250 N g_1^2 g_2^2 g_3^2 
\right. \nonumber \\
&& ~~~~~~~~~~~~~~~~ \left.
+~ 54000 N g_1^2 g_2^2 g_3 g_4 
- 1620 N^2 g_1^2 g_2^2 g_4^2 
+ 739500 N g_1^2 g_2^2 g_4^2 
\right. \nonumber \\
&& ~~~~~~~~~~~~~~~~ \left.
-~ 2625 N g_1^2 g_2 g_3^2 g_4 
+ 55500 N g_1^2 g_2 g_3 g_4^2 
- 964500 N g_1^2 g_2 g_4^3 
\right. \nonumber \\
&& ~~~~~~~~~~~~~~~~ \left.
-~ 810 N g_1^2 g_3^2 g_4^2 
- 10500 N g_1^2 g_3 g_4^3 
+ 249540 N g_1^2 g_4^4 
- 14250 N^2 g_2^4 g_4^2 
\right. \nonumber \\
&& ~~~~~~~~~~~~~~~~ \left.
+~ 31500 N g_2^4 g_4^2 
+ 157500 N g_2^3 g_3 g_4^2 
- 207000 N g_2^3 g_4^3 
- 37875 N g_2^2 g_3^2 g_4^2 
\right. \nonumber \\
&& ~~~~~~~~~~~~~~~~ \left.
+~ 168000 N g_2^2 g_3 g_4^3 
- 231000 N g_2^2 g_4^4 
+ 24000 g_3^4 g_4^2 
- 19500 g_3^3 g_4^3 
\right. \nonumber \\
&& ~~~~~~~~~~~~~~~~ \left.
-~ 929250 g_3^2 g_4^4 
+ 2820000 g_3 g_4^5 
+ 4089000 g_4^6 \right] ~+~ O ( g_i^8 ) \nonumber \\
\gamma_\phi(g_i) &=& 
\frac{1}{6} [3 \alpha g_1^2 - 10 g_1^2 + g_2^2] \nonumber \\
&& +~ \frac{1}{2160} \left[ - 196 N g_1^4 + 2750 g_1^4 + 420 g_1^2 g_2^2
- 4560 g_1^2 g_2 g_4 + 1200 g_1^2 g_4^2 - 110 N g_2^4
\right. \nonumber \\
&& \left. ~~~~~~~~~~~
+ 130 g_2^4 + 240 g_2^3 g_3 - 55 g_2^2 g_3^2 - 780 g_2^2 g_4^2 \right]
\nonumber \\
&& +~ \frac{1}{3888000} \left[  
2648 N^2 g_1^6 
+ 2592000 \zeta_3 N g_1^6 
- 5723500 N g_1^6 
+ 5832000 \zeta_3 g_1^6 
\right. \nonumber \\
&& ~~~~~~~~~~~~~~~~ \left.
+~ 2066000 g_1^6 
- 3888000 \zeta_3 N g_1^4 g_2^2 
+ 5255900 N g_1^4 g_2^2 
+ 4536000 \zeta_3 g_1^4 g_2^2 
\right. \nonumber \\
&& ~~~~~~~~~~~~~~~~ \left.
-~ 5595000 g_1^4 g_2^2 
- 2598000 N g_1^4 g_2 g_4 
+ 31806000 g_1^4 g_2 g_4 
\right. \nonumber \\
&& ~~~~~~~~~~~~~~~~ \left.
+~ 341700 N g_1^4 g_4^2 
+ 7776000 \zeta_3 g_1^4 g_4^2 
- 34440000 g_1^4 g_4^2 
\right. \nonumber \\
&& ~~~~~~~~~~~~~~~~ \left.
+~ 1296000 \zeta_3 N g_1^2 g_2^4 
- 3215000 N g_1^2 g_2^4 
- 1944000 \zeta_3 g_1^2 g_2^4 
\right. \nonumber \\
&& ~~~~~~~~~~~~~~~~ \left.
+~ 3086000 g_1^2 g_2^4 
+ 201000 g_1^2 g_2^3 g_3 
- 363000 N g_1^2 g_2^3 g_4 
+ 2856000 g_1^2 g_2^3 g_4 
\right. \nonumber \\
&& ~~~~~~~~~~~~~~~~ \left.
+~ 243750 g_1^2 g_2^2 g_3^2 
- 1476000 g_1^2 g_2^2 g_3 g_4 
- 1285500 N g_1^2 g_2^2 g_4^2 
\right. \nonumber \\
&& ~~~~~~~~~~~~~~~~ \left.
-~ 15552000 \zeta_3 g_1^2 g_2^2 g_4^2 
+ 8889000 g_1^2 g_2^2 g_4^2 
- 133500 g_1^2 g_2 g_3^2 g_4 
\right. \nonumber \\
&& ~~~~~~~~~~~~~~~~ \left.
+~ 7776000 \zeta_3 g_1^2 g_2 g_3 g_4^2 
- 13026000 g_1^2 g_2 g_3 g_4^2 
+ 15552000 \zeta_3 g_1^2 g_2 g_4^3 
\right. \nonumber \\
&& ~~~~~~~~~~~~~~~~ \left.
-~ 45186000 g_1^2 g_2 g_4^3 
- 258750 g_1^2 g_3^2 g_4^2 
+ 1710000 g_1^2 g_3 g_4^3 
\right. \nonumber \\
&& ~~~~~~~~~~~~~~~~ \left.
-~ 2925000 g_1^2 g_4^4 
- 6500 N^2 g_2^6 
+ 58000 N g_2^6 
- 648000 \zeta_3 g_2^6 
\right. \nonumber \\
&& ~~~~~~~~~~~~~~~~ \left.
+~ 1133000 g_2^6 
- 661500 N g_2^5 g_3 
+ 408000 g_2^5 g_3 
+ 96500 N g_2^4 g_3^2 
\right. \nonumber \\
&& ~~~~~~~~~~~~~~~~ \left.
-~ 648000 \zeta_3 g_2^4 g_3^2 
+ 1470250 g_2^4 g_3^2 
- 2172000 N g_2^4 g_4^2 
+ 879000 g_2^4 g_4^2 
\right. \nonumber \\
&& ~~~~~~~~~~~~~~~~ \left.
-~ 117750 g_2^3 g_3^3 
+ 63000 g_2^3 g_3 g_4^2 
- 9540000 g_2^3 g_4^3 
- 40875 g_2^2 g_3^4 
\right. \nonumber \\
&& ~~~~~~~~~~~~~~~~ \left.
+~ 123000 g_2^2 g_3^2 g_4^2 
- 2256000 g_2^2 g_3 g_4^3 
- 28626000 g_2^2 g_4^4 \right] ~+~ O ( g_i^8 ) \nonumber \\
\gamma_\sigma(g_i) &=& 
\frac{1}{12} [2 N g_2^2 + g_3^2 + 60 g_4^2] 
\nonumber \\
&& +~ \frac{1}{2160} \left[ 3820 N g_1^2 g_2^2 - 2400 N g_1^2 g_2 g_4 
+ 1656 N g_1^2 g_4^2 + 20 N g_2^4 + 480 N g_2^3 g_3 
\right. \nonumber \\
&& \left. ~~~~~~~~~~~
- 110 N g_2^2 g_3^2 + 65 g_3^4 - 780 g_3^2 g_4^2 + 12000 g_3 g_4^3 
+ 12960 g_4^4 \right] \nonumber \\
&& +~ \frac{1}{7776000} \left[ 
2570600 N^2 g_1^4 g_2^2 
+ 45360000 \zeta_3 N g_1^4 g_2^2 
+ 8144000 N g_1^4 g_2^2 
\right. \nonumber \\
&& ~~~~~~~~~~~~~~~~ \left.
-~ 849600 N^2 g_1^4 g_2 g_4 
- 118344000 N g_1^4 g_2 g_4 
+ 107856 N^2 g_1^4 g_4^2 
\right. \nonumber \\
&& ~~~~~~~~~~~~~~~~ \left.
-~ 85536000 \zeta_3 N g_1^4 g_4^2 
+ 190980000 N g_1^4 g_4^2 
+ 2592000 \zeta_3 N g_1^2 g_2^4 
\right. \nonumber \\
&& ~~~~~~~~~~~~~~~~ \left.
+~ 420000 N g_1^2 g_2^4 
+ 16584000 N g_1^2 g_2^3 g_3 
+ 18216000 N g_1^2 g_2^3 g_4 
\right. \nonumber \\
&& ~~~~~~~~~~~~~~~~ \left.
+~ 2592000 \zeta_3 N g_1^2 g_2^2 g_3^2 
- 7405000 N g_1^2 g_2^2 g_3^2 
+ 23328000 \zeta_3 N g_1^2 g_2^2 g_3 g_4 
\right. \nonumber \\
&& ~~~~~~~~~~~~~~~~ \left.
-~ 15336000 N g_1^2 g_2^2 g_3 g_4 
+ 62208000 \zeta_3 N g_1^2 g_2^2 g_4^2 
- 143424000 N g_1^2 g_2^2 g_4^2 
\right. \nonumber \\
&& ~~~~~~~~~~~~~~~~ \left.
-~ 192000 N g_1^2 g_2 g_3^2 g_4 
- 7824000 N g_1^2 g_2 g_3 g_4^2 
- 4608000 N g_1^2 g_2 g_4^3 
\right. \nonumber \\
&& ~~~~~~~~~~~~~~~~ \left.
-~ 1536000 N g_1^2 g_3^2 g_4^2 
+ 8079360 N g_1^2 g_3 g_4^3 
+ 5757840 N g_1^2 g_4^4 
\right. \nonumber \\
&& ~~~~~~~~~~~~~~~~ \left.
+~ 1381000 N^2 g_2^6 
- 1296000 \zeta_3 N g_2^6 
+ 2140000 N g_2^6 
- 576000 N^2 g_2^5 g_3 
\right. \nonumber \\
&& ~~~~~~~~~~~~~~~~ \left.
-~ 264000 N g_2^5 g_3 
- 1500 N^2 g_2^4 g_3^2 
- 3240000 \zeta_3 N g_2^4 g_3^2 
+ 6661500 N g_2^4 g_3^2 
\right. \nonumber \\
&& ~~~~~~~~~~~~~~~~ \left.
+~ 4530000 N g_2^4 g_4^2 
+ 390000 N g_2^3 g_3^3 
+ 864000 N g_2^3 g_3 g_4^2 
\right. \nonumber \\
&& ~~~~~~~~~~~~~~~~ \left.
+~ 45216000 N g_2^3 g_4^3 
- 238000 N g_2^2 g_3^4 
- 5436000 N g_2^2 g_3^2 g_4^2 
\right. \nonumber \\
&& ~~~~~~~~~~~~~~~~ \left.
-~ 11664000 N g_2^2 g_3 g_4^3 
- 11838000 N g_2^2 g_4^4 
- 324000 \zeta_3 g_3^6 
+ 649375 g_3^6 
\right. \nonumber \\
&& ~~~~~~~~~~~~~~~~ \left.
-~ 21000 g_3^4 g_4^2 
- 6816000 g_3^3 g_4^3 
- 77760000 \zeta_3 g_3^2 g_4^4 
+ 110151000 g_3^2 g_4^4 
\right. \nonumber \\
&& ~~~~~~~~~~~~~~~~ \left.
+~ 154968000 g_3 g_4^5 
- 59460000 g_4^6 \right] 
~+~ O ( g_i^8 ) 
\label{anomdim3}
\end{eqnarray}
where $g_i$ denotes each of the possible four coupling constants, $\zeta_z$ 
is the Riemann zeta function and the order symbol represents all combinations 
of the couplings at that order. Also all our results are given in the $\MSbar$ 
scheme with the scheme dependence first arising at two loops in all these 
expressions including the $\beta$-functions since (\ref{lag6}) has more than 
one coupling constant. Next the $\beta$-functions are 
\begin{eqnarray}
\beta_1(g_i) &=& 
-~ \frac{1}{30} N g_1^3 \nonumber \\
&& +~ \frac{g_1}{1080} \left[ 10 N g_1^2 g_2^2 - 370 N g_1^4 
+ 6 N g_1^2 g_4^2 + 30 N g_2^2 g_4^2 + 15 g_3^2 g_4^2 - 120 g_3 g_4^3 
+ 660 g_4^4 \right] \nonumber \\
&& +~ O ( g_i^7 ) \nonumber \\
\beta_2(g_i) &=& 
\frac{1}{12} \left[ - 40 g_1^2 g_2 + 120 g_1^2 g_4 + 2 N g_2^3 - 8 g_2^3 
- 12 g_2^2 g_3 + g_2 g_3^2 + 60 g_2 g_4^2 \right] 
\nonumber \\
&& +~ \frac{1}{2160} \left[ 2832 N g_1^4 g_4 - 6032 N g_1^4 g_2 
- 13400 g_1^4 g_2 - 31200 g_1^4 g_4 + 3820 N g_1^2 g_2^3 
\right. \nonumber \\
&& \left. ~~~~~~~~~~~
- 6480 g_1^2 g_2^3 - 960 g_1^2 g_2^2 g_3 - 2400 N g_1^2 g_2^2 g_4 
- 36240 g_1^2 g_2^2 g_4 
\right. \nonumber \\
&& \left. ~~~~~~~~~~~
- 19440 g_1^2 g_2 g_3 g_4 + 1656 N g_1^2 g_2 g_4^2 
+ 88800 g_1^2 g_2 g_4^2 + 18480 g_1^2 g_3 g_4^2 
\right. \nonumber \\
&& \left. ~~~~~~~~~~~
+ 8160 g_1^2 g_4^3 - 860 N g_2^5 - 2680 g_2^5 + 1320 N g_2^4 g_3 
- 1800 g_2^4 g_3 - 110 N g_2^3 g_3^2 
\right. \nonumber \\
&& \left. ~~~~~~~~~~~
- 3140 g_2^3 g_3^2 - 4080 g_2^3 g_4^2 - 120 g_2^2 g_3^3 
- 5040 g_2^2 g_3 g_4^2 + 38880 g_2^2 g_4^3 + 65 g_2 g_3^4 
\right. \nonumber \\
&& \left. ~~~~~~~~~~~
- 780 g_2 g_3^2 g_4^2 + 12000 g_2 g_3 g_4^3 + 12960 g_2 g_4^4 \right] ~+~ 
O ( g_i^7 ) \nonumber \\
\beta_3(g_i) &=& 
\frac{1}{4} \left[ - 8 N g_2^3 + 2 N g_2^2 g_3 - 3 g_3^3 + 60 g_3 g_4^2 
- 160 g_4^3 \right]
\nonumber \\
&& +~ \frac{1}{720} \left[ 3820 N g_1^2 g_2^2 g_3 - 12720 N g_1^2 g_2^3 
- 19440 N g_1^2 g_2^2 g_4 - 2400 N g_1^2 g_2 g_3 g_4 
\right. \nonumber \\
&& \left. ~~~~~~~~~
+ 22560 N g_1^2 g_2 g_4^2 + 1656 N g_1^2 g_3 g_4^2 - 5664 N g_1^2 g_4^3 
- 240 N g_2^5 - 3220 N g_2^4 g_3
\right. \nonumber \\
&& \left. ~~~~~~~~~
- 600 N g_2^3 g_3^2 
+ 310 N g_2^2 g_3^3 - 625 g_3^5 - 3300 g_3^3 g_4^2 + 50880 g_3^2 g_4^3 
- 104160 g_3 g_4^4 
\right. \nonumber \\
&& \left. ~~~~~~~~~
- 61440 g_4^5 \right] ~+~ O ( g_i^7 ) \nonumber \\
\beta_4(g_i) &=& 
\frac{1}{60} \left[ 20 N g_1^2 g_2 - 4 N g_1^2 g_4 + 10 N g_2^2 g_4 
+ 5 g_3^2 g_4 - 40 g_3 g_4^2 + 220 g_4^3 \right] 
\nonumber \\
&& +~ \frac{1}{10800} \left[ 37400 N g_1^4 g_2 + 400 N g_1^4 g_4 
- 200 N g_1^2 g_2^3 + 2100 N g_1^2 g_2^2 g_3 
\right. \nonumber \\
&& \left. ~~~~~~~~~~~~
+ 33700 N g_1^2 g_2^2 g_4 
- 400 N g_1^2 g_2 g_3 g_4 - 10800 N g_1^2 g_2 g_4^2 
- 296 N g_1^2 g_3 g_4^2 
\right. \nonumber \\
&& \left. ~~~~~~~~~~~~
+ 7216 N g_1^2 g_4^3 + 100 N g_2^4 g_4 
+ 2400 N g_2^3 g_3 g_4 - 1200 N g_2^3 g_4^2 
\right. \nonumber \\
&& \left. ~~~~~~~~~~~~
- 550 N g_2^2 g_3^2 g_4 
+ 2000 N g_2^2 g_3 g_4^2 + 2600 N g_2^2 g_4^3 + 325 g_3^4 g_4 
\right. \nonumber \\
&& \left. ~~~~~~~~~~~~
+ 400 g_3^3 g_4^2 - 13800 g_3^2 g_4^3 + 4000 g_3 g_4^4 + 205200 g_4^5 \right] 
~+~ O ( g_i^7 ) ~. 
\label{beta2}
\end{eqnarray}
These two loop $\beta$-functions complete the renormalization of (\ref{lag6})
to this order. In terms of being confident that the results are correct we note
that we have implemented the automatic renormalization algorithm of \cite{36}. 
In other words we evaluate all the contributing graphs in terms of the bare 
parameters which are the coupling constants and gauge parameter. Then their 
renormalized counterparts are introduced by a multiplicative rescaling without 
having to follow the method of subtractions. This means that all the double 
poles of the two loop renormalization constants are already determined by their
one loop simple poles and therefore we have verified that these correctly 
emerge. By the same token we have been able to check the two loop 
$\beta$-functions by computing the field anomalous dimensions to {\em three} 
loops. In this case double and triple poles of the three loop renormalization 
constants are fixed by lower loop information including the two loop coupling 
constant renormalization constants. Again we confirm that the results of 
(\ref{anomdim3}) are consistent with this check. This is also the reason for 
the large expressions in (\ref{anomdim3}) compared with (\ref{beta2}). For 
completeness we note that to determine the anomalous dimensions 
(\ref{anomdim3}) the number of three loop graphs computed were $561$, $428$ and
$572$ for the $A_\mu$, $\sigma$ and $\phi^i$ $2$-point functions respectively. 
Another independent check on our computations rests in the Ward-Takahashi 
identity associated with the $U(1)$ gauge field. As in QED the gauge field 
anomalous dimension is not independent and is related to the gauge 
$\beta$-function. In other words this identity implies 
\begin{equation}
\beta_1(g_i) ~=~ g_1 \gamma_A(g_i)  
\label{wti}
\end{equation}
in our notation here and we note that it is clearly satisfied to two loops in 
the $\MSbar$ scheme from comparing (\ref{anomdim3}) and (\ref{beta2}). By the 
same token we now know $\beta_1(g_i)$ to three loops but the remaining 
$\beta$-functions need to be computed explicitly which is beyond the scope of 
this article.

As a final comment on our renormalization group functions we note that we 
carried out our computations in an arbitrary linear covariant gauge. While in 
the $\MSbar$ scheme this means that the $\beta$-functions do not depend on the 
gauge parameter $\alpha$, the anomalous dimensions are in fact gauge parameter 
dependent. However in (\ref{anomdim3}) the only place where $\alpha$ appears is
in the one loop term of the $\phi^i$ anomalous dimension. We note that in 
\cite{23} the one loop computations were performed solely in the Landau gauge. 
Although this dependence on $\alpha$ may appear to be peculiar by contrast it 
now seems to be a feature of any $U(1)$ gauge theory in the $\MSbar$ scheme,
independent of dimension, since the same property is present in four 
dimensional QED from explicit computations, 
\cite{37,38,39,40,41,42,43,44,45,46,47}, as well as in higher dimensional 
versions of QED, \cite{48,49,29}. An interesting and novel insight into 
understanding the underlying reasons for this property using graphical methods 
that transcends the spacetime dimension has been developed in 
\cite{50,51,52,53}. 

While these represent the main internal checks on any perturbative multiloop
renormalization one also has to connect with the underlying universal theory 
that (\ref{lag4}) and (\ref{lag6}) are partners to. To achieve this we note 
that information on the universal structure is accessed through the 
$d$-dependent critical exponents that define the properties of the 
Wilson-Fisher fixed point and are renormalization group invariants. These can 
be deduced through the large $N$ expansion approach of \cite{6,7,8} where $1/N$
acts as a dimensionless coupling constant in $d$-dimensions. Expanding the 
exponents in an $\epsilon$ expansion near the critical dimension of the theory 
then they will be in one-to-one correspondence with the large $N$ and 
$\epsilon$ expansion of (\ref{anomdim3}) at the Wilson-Fisher fixed point. 
Therefore we now record the details of this exercise but first recall that 
(\ref{lag6}) contains two main universality classes of interest depending on 
which of the fields $A_\mu$ and $\sigma$ are active, \cite{23}. One corresponds
to the full $\Cc\Pp(N)$ class when both are present. When only $A_\mu$ is 
active then one is in the scalar QED universality class which provides us with 
another set of exponents to compare with available large $N$ exponents. 
Strictly there is a third universality class in (\ref{lag6}) when $A_\mu$ is 
inactive. This corresponds to a complexified scalar and lies in the same 
universality class as the $O(2N)$ nonlinear $\sigma$ which also contains 
$\phi^4$ theory in four dimensions as well as six dimensional $\phi^3$ theory
studied in \cite{1,19,20,21,22}. However we will not present any connections 
here for this fixed point since the corresponding large $N$ analysis has been 
given elsewhere \cite{1,22}. Instead we merely note that when $g_1$ and $g_4$ 
are set to zero the same renormalization group functions for six dimensional 
$O(2N)$ $\phi^3$ theory emerge consistent with \cite{1,19,20,21,22}. 

First we concentrate on the full $\Cc\Pp(N)$ universality class represented by
(\ref{lag6}) in six dimensions and note that at the Wilson-Fisher fixed point
the critical couplings are 
\begin{eqnarray}
g_1^\ast &=& i \sqrt{\frac{30\epsilon}{N}} 
\left[ \frac{1}{N} ~+~ 155 \frac{\epsilon}{N^2} ~+~ 
O \left( \epsilon^2; \frac{1}{N^3} \right) \right] \nonumber \\
g_2^\ast &=& -~ \sqrt{\frac{30\epsilon}{N}} 
\left[ \frac{1}{5N} ~+~ \frac{336}{5N^2} ~-~ 67 \frac{\epsilon}{N^2} ~+~
O \left( \epsilon^2; \frac{1}{N^3} \right) \right] \nonumber \\
g_3^\ast &=& -~ \sqrt{\frac{30\epsilon}{N}} 
\left[ \frac{6}{5N} ~+~ \frac{8736}{5N^2} ~-~ 
1494 \frac{\epsilon}{N^2} ~+~ 
O \left( \epsilon^2; \frac{1}{N^3} \right) \right] \nonumber \\
g_4^\ast &=& -~ \sqrt{\frac{30\epsilon}{N}} 
\left[ \frac{1}{N} ~-~ \frac{224}{N^2} ~+~ 743 \frac{\epsilon}{N^2} ~+~ 
O \left( \epsilon^2; \frac{1}{N^3} \right) \right]
\label{cpncc}
\end{eqnarray}
where the leading orders agree with those of \cite{23}. Given these we find
\begin{eqnarray}
\gamma_\phi(g_i^\ast) &=& \left[ 51 \epsilon ~-~ \frac{167}{2} \epsilon^2 ~+~
O(\epsilon^3) \right] \frac{1}{N} ~+~ O \left( \frac{1}{N^2} \right) 
\nonumber \\
\gamma_\sigma(g_i^\ast) &=& \epsilon ~+~ 
\left[ 1440 \epsilon ~-~ 3456 \epsilon^2 ~+~
O(\epsilon^3) \right] ~+~ O \left( \frac{1}{N^2} \right) \nonumber \\
\gamma_A(g_i^\ast) &=& \epsilon ~+~ 
O \left( \epsilon^3; \frac{1}{N^2} \right)
\end{eqnarray}
at leading order in large $N$ in the Landau gauge. The absence of $O(1/N)$ 
corrections for the gauge field dimension derives from the way the universal
theory (\ref{lag2}) is formulated and the Ward-Takahashi identity. In
particular the coupling constant in the $3$-point interaction of the gauge 
field with the matter field $\phi^i$ is absent in the definition of the 
underlying universal theory as is clear in (\ref{lag2}), (\ref{lag4}) or 
(\ref{lag6}). Consequently in the critical point approach used in \cite{26} the
gauge field has no anomalous dimension. This is similar to what has been 
observed in the large $N$ expansion of other abelian gauge theories. Expanding 
the $d$-dimensional expressions for the Landau gauge large $N$ exponents of the 
universal theory, \cite{26}, in $d$~$=$~$6$~$-$~$2\epsilon$ we find exact 
agreement. The reason why the checks are carried out in the Landau gauge is 
that the gauge parameter in effect acts as a second coupling constant. 
Therefore since we are considering a fixed point one has to find the critical 
value of the gauge parameter akin to finding (\ref{cpncc}). In the case of 
$\alpha$ its critical value is zero.

Before repeating the same exercise for the scalar QED universality class we 
note that the {\em three} loop $\MSbar$ renormalization group functions are
\begin{eqnarray}
\gamma_\psi(g_1) &=& \frac{[3 \alpha - 10]}{6} g_1^2 
- \frac{[98 N - 1375]}{1080} g_1^4 \nonumber \\
&& +~ [662 N^2 + 648000 N \zeta_3 - 1430875 N + 1458000 \zeta_3 + 516500 ]
\frac{g_1^6}{972000} ~+~ O(g_1^8) \nonumber \\
\gamma_A(g_1) &=& -~ \frac{N}{30} g_1^2 - \frac{37N}{108} g_1^4
- \frac{N}{48600} [ 1017 N + 59600 ] g_1^6 ~+~ O(g_1^8) \nonumber \\
\beta_1(g_1) &=& -~ \frac{N}{30} g_1^3 - \frac{37N}{108} g_1^5
- \frac{N}{48600} [ 1017 N + 59600 ] g_1^7 ~+~ O(g_1^9) 
\end{eqnarray}
where $g_1$ is the only active coupling and we have used the Ward-Takahashi 
identity (\ref{wti}) to deduce $\beta_1(g_1)$ here. Therefore expanding the 
fixed point in large $N$ from (\ref{beta2}) we have 
\begin{eqnarray}
g_1^\ast &=& i \sqrt{\frac{30\epsilon}{N}} 
\left[ \frac{1}{N} ~+~ \frac{925\epsilon}{6N^2} ~-~
\frac{565\epsilon^2}{2N^2} ~+~ O \left( \epsilon^2; \frac{1}{N^3} \right) 
\right] \nonumber \\
g_2^\ast &=& g_3^\ast ~=~ g_4^\ast ~=~ O \left( \epsilon; \frac{1}{N^3}\right)
\end{eqnarray}
and find 
\begin{eqnarray}
\gamma_\phi(g_i^\ast) &=& \left[ 50 \epsilon ~-~ \frac{245}{3} \epsilon^2 ~-~
\frac{331}{18} \epsilon^3 ~+~ O(\epsilon^4) \right] \frac{1}{N} ~+~ 
O \left( \frac{1}{N^2} \right) \nonumber \\
\gamma_A(g_i^\ast) &=& \epsilon ~+~ 
O \left( \epsilon^4; \frac{1}{N^2} \right) ~.
\end{eqnarray}
These are also in agreement with the expression for the field critical 
exponents also available in \cite{26} and again the gauge field has no large 
$N$ corrections. 

\sect{Discussion.}

We have extended the one loop analysis of \cite{23,24} to two loops and
established the ultraviolet completion of the six dimensional $\Cc\Pp(N)$ 
$\sigma$ model which also contains the scalar QED universality class as a 
sub-theory to this new order. As a gauge theory it shares similar features to 
the non-abelian gauge theory in six dimensions studied in \cite{29,30}. For 
instance at one loop the gauge $\beta$-function depends only on the gauge 
coupling and moreover like six dimensional QED the gauge coupling is 
asymptotically free as was shown in \cite{48,49}. That the same feature emerges
in the scalar case is a consequence of the underlying gauge symmetry. Indeed 
there are other general structural similarities with four dimensional QED. One 
of these is that the gauge parameter is only present at one loop and not two 
loops in the $\phi^i$ field anomalous dimension. Not only is this feature 
present in QED but it would appear that the graphical proof of this given in 
\cite{52,53} for QED could be simply adapted to show this to all orders in 
perturbation theory. In terms of other future work in this universality class 
one thing that is lacking is higher order large $N$ critical exponents for both
the field dimensions and the critical $\beta$-function slopes. This would 
require the extension of the original formalism developed in \cite{6,7} for the
$O(N)$ $\phi^4$ universality class that produced $O(1/N^2)$ and $O(1/N^3)$ 
exponents in $d$-dimensions. Given that $O(1/N^2)$ exponents are available for 
QED, \cite{54,55,56}, the application to the $\Cc\Pp(N)$ case ought not to be 
problematic. From another direction the next theory in the tower of the 
universality class will become active in eight dimensions. It should have a 
Lagrangian of the form
\begin{eqnarray}
L^{(8)} &=& \overline{D_\mu \phi}^{\,i} D^\mu \phi^i ~+~ 
\frac{1}{2} \left( \Box \sigma \right)^2 ~-~ 
\frac{1}{4} \left( \partial_\mu \partial_\nu F_{\sigma\rho} \right)
\left( \partial^\mu \partial^\nu F^{\sigma\rho} \right) ~-~ 
\frac{1}{2\alpha} \left( \Box \partial^\mu A_\mu \right)
\left( \Box \partial^\nu A_\nu \right) \nonumber \\
&& +~ g_2 \sigma \bar{\phi}^i \phi^i ~+~ 
\frac{g_3}{6} \sigma^2 \Box \sigma ~+~
\frac{g_4}{2} \left( \Box \sigma \right) \! F_{\mu\nu} F^{\mu\nu} ~+~
g_5 \sigma F_{\mu\nu} \Box F^{\mu\nu} ~+~
\frac{g_6^2}{24} \sigma^4 \nonumber \\
&& +~ \frac{g_7^2}{32} F_{\mu\nu} F^{\mu\nu} F_{\sigma\rho} F^{\sigma\rho} ~+~
\frac{g_8^2}{8} F_{\mu\nu} F^{\mu\sigma} F_{\nu\rho} F^{\sigma\rho} ~+~
\frac{g_9^2}{4} \sigma^2 F_{\mu\nu} F^{\mu\nu} 
\label{lag8}
\end{eqnarray}
which includes a different set of what are termed spectator interactions that 
are independent. The theory relevant for the scalar QED universality class 
involves the spectator couplings $g_7$ and $g_8$, in addition to $g_1$. The two
associated operators are also present in the eight dimensional version of QED, 
\cite{29}. To repeat the two loop analysis carried out here for (\ref{lag8}) is
beyond the scope of the present article. However having information on its 
renormalization group functions would additionally complement any future 
determination of the $d$-dimensional $O(1/N^2)$ critical exponents.

\vspace{1cm}
\noindent
{\bf Acknowledgements.} This work was supported by a DFG Mercator Fellowship.
The author thanks Dr H. Khachatryan for several valuable discussions.

\end{document}